\documentclass[journal]{vgtc} 

\usepackage{cite}
\usepackage[bottom]{footmisc}
\usepackage{mathptmx}
\usepackage{graphicx}
\usepackage{siunitx}
\usepackage{times}
\usepackage[T1]{fontenc}
\usepackage[font=normalsize]{subcaption}
\usepackage{enumitem}
\usepackage[table]{xcolor} 
\usepackage{wrapfig}
\usepackage{multirow}
\usepackage{adjustbox}
\usepackage{mwe}
\usepackage{graphbox}
\usepackage{dirtytalk}
\usepackage{soul}
\usepackage{float}
\usepackage{mdframed}
\usepackage{tabularx}
\usepackage{xcolor}
\usepackage[linesnumbered,ruled,vlined]{algorithm2e}

\definecolor{greenB}{RGB}{77, 175, 74}
\definecolor{purpleF}{RGB}{152,78,163}
\definecolor{CindySalmon}{RGB}{232, 125, 114}

\newcommand*\inline[1]{\includegraphics[height=.9em]{#1}
}
\newcommand{\icon}[1]{\inline{image/icons/#1.pdf}}
\newcommand{\revisiondelete}[1]{\relax}
\newcommand{\revision}[1]{\textcolor{black}{#1}}

\PassOptionsToPackage{hyphens}{url}
\usepackage[]{hyperref} 
\hypersetup{
 linkcolor=black,
 colorlinks=true,
 linkcolor= black,
 citecolor= black,
 pageanchor=true,
 urlcolor = black,
 plainpages = false,
 pdfauthor = {},
 pdftitle = {},
 pdfsubject = {},
 pdfkeywords = {},
 linktocpage
}

\onlineid{1151} 
\vgtccategory{Theoretical and Empirical} 
\vgtcinsertpkg 

\title{Motion-Based Visual Encoding Can Improve Performance on \\Perceptual Tasks with Dynamic Time Series}

\author{Songwen Hu, Ouxun Jiang, Jeffrey Riedmiller, Cindy Xiong Bearfield}
\authorfooter{ 
\item
Songwen Hu and Cindy Xiong Bearfield are with Georgia Tech.
\newline
E-mail: cxiong, shu343@gatech.edu
\item
Ouxun Jiang is with Northwestern University.
\item
Jeffrey Riedmiller is with Dolby Laboratories
E-mail: jcr@dolby.com
}

\shortauthortitle{Xiong \MakeLowercase{\textit{et al.}}: }
\teaser{
  \centering
  \setlength{\abovecaptionskip}{0pt}
  \includegraphics[width=0.85\linewidth, alt={Examples of different animation design options.}]{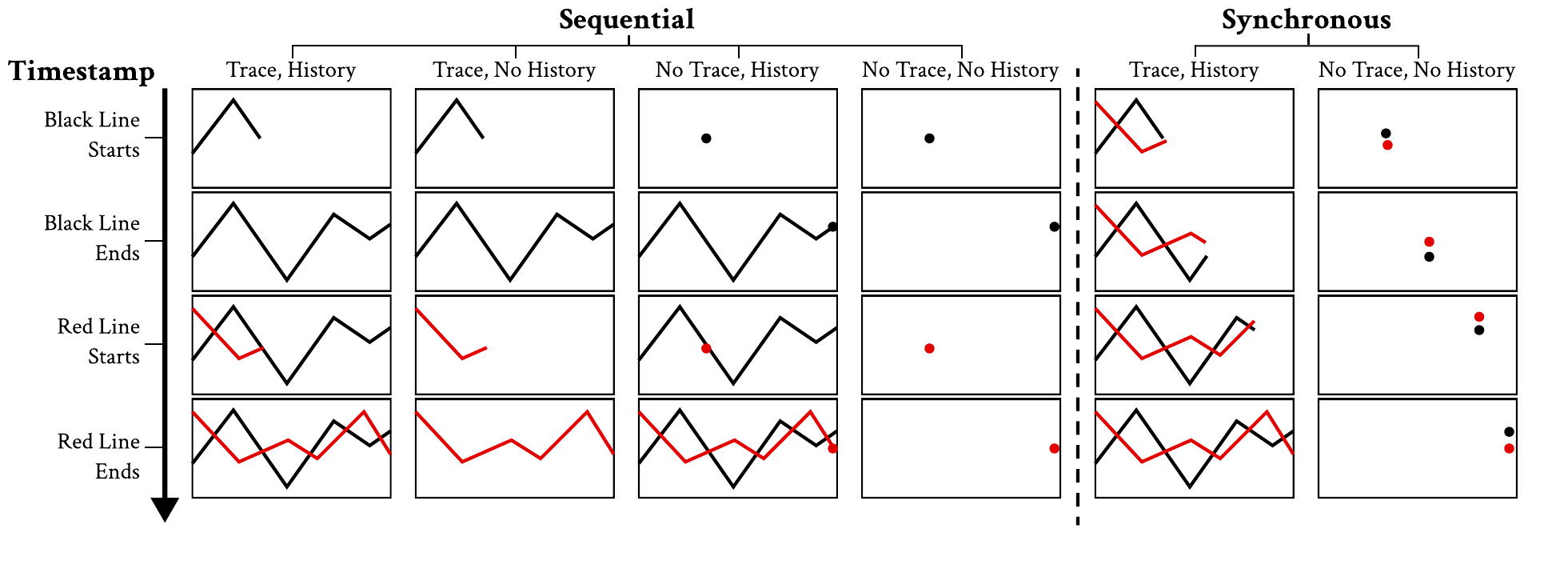}
  \vspace{-1em}
  \caption{\revision{This figure illustrates different animation design options for animated line charts. The animations are arranged in a time sequence from top to bottom and categorized into six conditions from left to right.}}
  \label{fig:teaser}
}
\abstract{
Dynamic data visualizations can convey large amounts of information over time, such as using motion to depict changes in data values for multiple entities. Such dynamic displays put a demand on our visual processing capacities, yet our perception of motion is limited. 
Several techniques have been shown to improve the processing of dynamic displays. Staging the animation to sequentially show steps in a transition and tracing object movement by displaying trajectory histories can improve processing by reducing the cognitive load. In this paper, We examine the effectiveness of staging and tracing in dynamic displays. We showed participants animated line charts depicting the movements of lines and asked them to identify the line with the highest mean and variance. We manipulated the animation to display the lines with or without staging, tracing and history, and compared the results to a static chart as a control. Results showed that tracing and staging are preferred by participants, and improve their performance in mean and variance tasks respectively. They also preferred display time 3 times shorter when staging is used. Also, encoding animation speed with mean and variance in congruent tasks is associated with higher accuracy. These findings help inform real-world best practices for building dynamic displays. The supplementary materials can be found at \url{https://osf.io/8c95v/}
} 

\keywords{Animation, Dynamic Displays, Perception, Motion, Analytic Tasks}

\begin{document}
\maketitle
\section{Introduction}
Visualizations leverage the strength of people's perceptual abilities to enhance data analysis. 
Dynamic visualizations, such as multi-class animated time series that show changes in a large volume of data over time, have become increasingly popular with technological advances in visualization tools~\cite{robertson2008effectiveness, tekusova2007applying, heer2007animated}.
While visualization designers and data journalists benefit from the ability of dynamic visualizations to engage audiences, the large volumes of data presented can be effortful for the viewers to process. 
Researchers criticize the transient nature of dynamic displays, as data tends to overlap and move too swiftly across the display, decreasing the legibility of the key statistics~\cite{tversky2002animation, hegarty2004dynamic}. 
However, making sense of dynamic data can be a necessity in many scenarios, from health care to network services~\cite{fayn1996interactive, netflix}.
For example, medical professionals might need to quickly extract summary statistics from ECG data and monitor anomalies, and network engineers might need to continuously compare network stability to identify unusual fluctuations to optimize data streaming. 
So how can we make dynamic displays more effective?

In this paper, we explore visual manipulation techniques that might improve viewer performance in perceptual tasks with dynamic displays, such as extracting summary statistics or identifying anomalies. 
Our work focuses on understanding human perception of \textit{dynamic time series}, where data values are “streamed” into the chart display space following a time variable. 
This supplements the traditionally more well-studied animated displays used for transitions and interactive systems in the visualization field~\cite{heer2012interactive}, such as transitions between different static charts representing the same dataset~\cite{heer2007animated}, interactive transitions to afford exploration of radial graphs~\cite{yee2001animated}, and transitions between views of a scatterplot matrix~\cite{rodrigues2024comparative}.
Visual manipulation techniques used in these animated transitions, such as “staging” and “tracing”, are implemented differently than in dynamic time series.
We define "staging" in dynamic time series as sequentially, rather than synchronously presenting data.
In a time series with two lines, a "staged" version would display one line and then the next line. 
This contrasts with the notion of "staging" in animated transitions, which focuses on showing different phases of a transition by breaking it up into a set of simple sub-transitions~\cite{heer2007animated}. 
In dynamic time series, we define "tracing" as a process similar to animated transitions, where the path of an object's movement is displayed.
Imagine a line in a time series moving from left to right through time.
With "tracing", the line would grow from left to right as more data is displayed on the screen.
Without "tracing", it would become a single dot moving from left to right.
We added an additional "history" dimension for when the data values are sequentially displayed (staged)~\cite{horowitz2007tracking, fisher2010anim}.
The data "history" is considered preserved when the movement trajectory of the previous object remains on the chart even after the subsequent object starts moving, and it is not preserved when the movement trajectory of a previous objective gets erased when the subsequent object starts to move. 
A staged dynamic time series will keep the trace of the first line on the chart when displaying the second line if it is "preserving history", otherwise it would erase the trace of the first line when displaying the second line. 
Figure~\ref{fig:teaser} shows combinations of staging, tracing, and history preservation. 

We quantify the effects of these visual manipulations---staging, tracing, and history preservation---in dynamic time series, on perceptual analysis tasks, and compare user performance to that in static displays.
This will provide visualization designers with a better understanding of the trade-offs between using animation for engagement and using static visualizations for higher visual precision and less effortful perceptual analysis~\cite{hegarty2004dynamic}. 

Generally, we found dynamic displays to facilitate mean and variance comparison tasks, but not outlier detection tasks. 
In addition to objective performance measures on perceptual analysis tasks, we also considered people's subjective preferences, diving into what people \textit{can} take away from dynamic displays with qualitative data. The most preferred condition was static visualizations, and people prefer animations with history over those without. Overall, animations with traces are preferred over those without, unless the animation without traces includes history. Participants' preference for the visualizations, however, does not match their performances. We conducted a Kendall's Tau correlation test for the accuracy rankings for the perceptual tasks and for the preference ranking when there were four lines (preference only measured for four lines). All correlations were low: the preference ranking's correlation was -0.05 with the mean accuracy ranking, 0.24 with variance accuracy ranking, and 0.05 with the significant outlier accuracy ranking.

To further improve perceptual task performance with dynamic time series, we also explore another venue of visual manipulation: motion-based data encoding. 
Seminal work in visualization perception by Cleveland and McGill~\cite{cleveland1984graphical}, Munzner~\cite{munzner2014visualization}, and MacKinlay~\cite{mackinlay1986automating} identified multiple visual encoding channels, such as color, size, and orientation, and provided guidelines on the effectiveness of encoding channels depending on the data type. 
For example, size encoding can effectively show an increase in quantity but not in uncertainty~\cite{liu2018visualizing}.
More recent work has demonstrated that redundantly encoding data values with multiple visual channels can increase perceptual efficiency when interpreting visualized data~\cite{nothelfer2017redundant}.
But despite the increasing popularity of animation and dynamic displays, the visualization community has yet to systematically explore the effectiveness of motion-based channels to supplement existing rankings of visual channels.
Motion is a powerful perceptual cue that tends to draw viewers' attention regardless of context relevance~\cite{hillstrom1994visual}. 
Its transient nature might help reduce clutter in a visualization depicting multiple data series. 
Visualization readers might have certain mental schema associated with motion that allows them to more intuitively interpret some data dimensions when they are encoded with motion, similar to semantic alignments in other visual channels such as color.
For example, a data item that is encoded with a color that aligns with its semantic is more easily interpreted, such as green with celery~\cite{schloss2023color}, or higher quantity with a darker color~\cite{schloss2018mapping}. 
Therefore, we posit that leveraging motion as an encoding channel can help analysts compare changing means, detect data fluctuations, and identify data points of potential concerns (e.g., outliers) more effectively. 
In this paper, we systematically investigate the effectiveness of redundantly encoding motion (with a focus on animation speed) with y-axis position, data variance, and outlier frequency.
We also point to future opportunities to consider combining the power of motion and redundant encoding to enhance visualizations.

\vspace{1mm}
\noindent \textbf{Contribution:} We contribute three experiments examining the effectiveness of animation techniques, including staging and tracing, on low-level analytic tasks with dynamic time series. 
Using six animated time series designs (plus a static version as the control), we compare objective measures such as readers' accuracy and speed at extracting mean and variance, and subjective measures such as reader preferences and takeaways from these animations.
The results allowed us to generate design guidelines for animated time series.
We also contribute a close investigation of \textit{animation speed} as a data encoding channel, identifying its potential strengths and shortcomings. 
\vspace{1mm}

\begin{figure}[!t]
    \centering
    \includegraphics[width=\columnwidth]{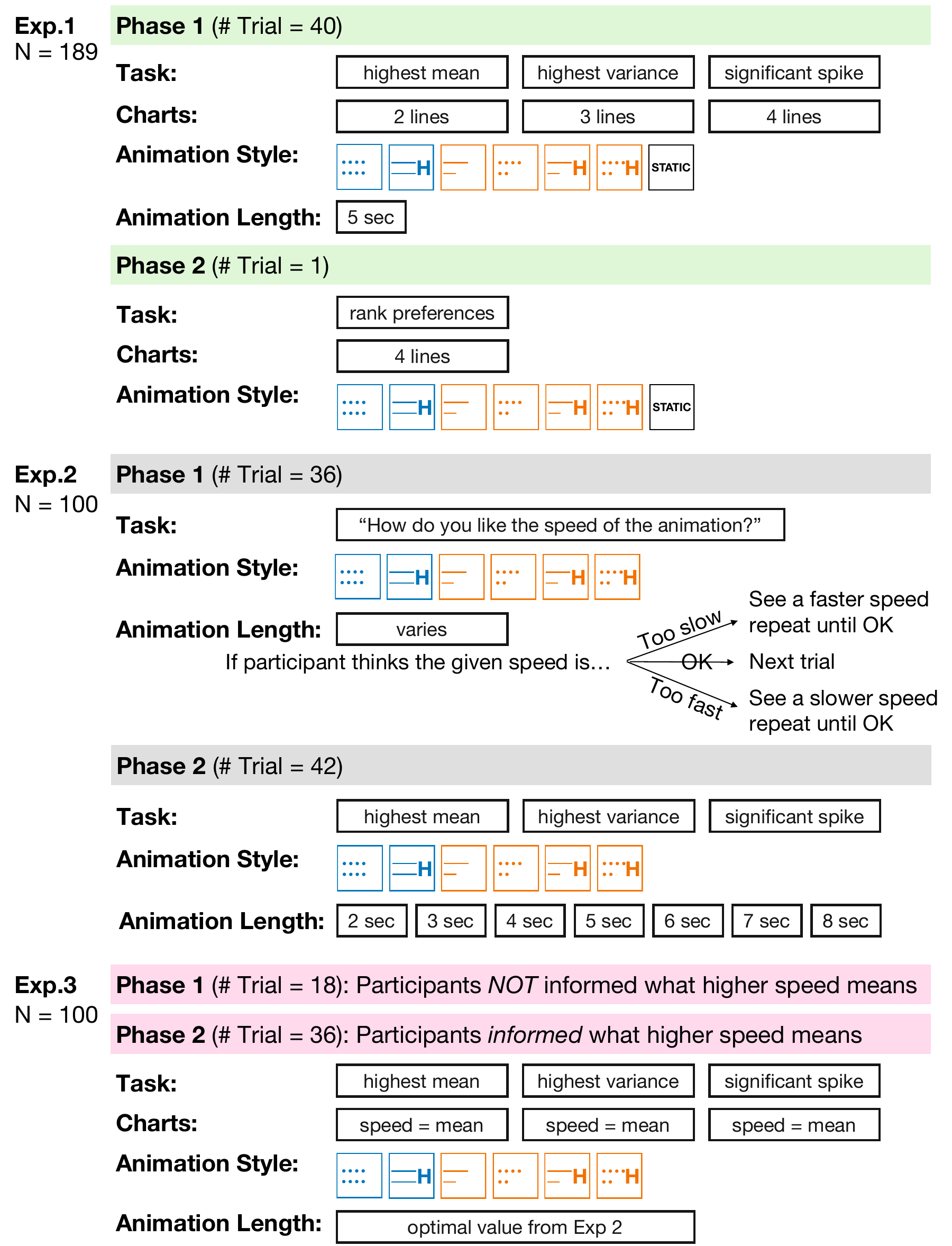}
     \caption{Set-up for Experiments 1-3.}
     \vspace{-1em}
    \label{fig:procedures}
\end{figure}

\noindent \textbf{Study Overview} \label{overview}
As outlined in Figure~\ref{fig:procedures}, Experiment 1 examines people's ability to extract summary statistics, such as mean and variance, and identify significant outliers from the dynamic time series across different amounts of data and animation designs.
Experiment 2 measures peoples' preferences for animation speed.
Experiment 3 investigates which data attribute is the most effectively encoded using speed, specifically focused on the average y-axis position, the variance, and the presence of outliers. 
Experiment 4 qualitatively examines conclusions people generate across the animated versus static displays.

\section{Related Work}
\label{section:rw}

\subsection{Visual Perception of Motion}
Our visual system has evolved to process a dynamic world. We can rapidly summarize the average speeds, directions, and trajectories of moving objects at high accuracy \cite{haberman2012ensemble, watamaniuk1992human, williams1984coherent}. It has been proved that visualizations in motion have perceptional benefits in some real-world applications, like a particle flow map\cite{ding2023data} and an algorithm \cite{ehlschlaeger1997visualizing} that generates intermediate frames in an animated visualization to show uncertainty.
In the literature review of animated visualization of time-oriented data by Kriglstein et.al \cite{kriglstein2012animation},
animated visualizations are used to "track the changes in data" and traces are used to "support users in analyzing trends in data or data that evolves over time."
When being presented with animations, people often find it engaging and efficient as animations can quickly depict the changes. But when being asked to explore or analyze data with animations, animation becomes slower and less accurate \cite{fisher2010anim}. In addition, animation is good at showing changes in data and works better for small-batch data (otherwise overlapping).
According to Robertson et al., although trend animations are the fastest and most engaging presentation technique, \textbf{participants were faster to analyze static trends with traces} than with animations which showed no path information \cite{robertson2008effectiveness}. Kriglstein found evidence that \textbf{participants are confused about traces, especially when more datasets are presented}\cite{kriglstein2012animation}. However, these findings lack systematic and quantitative models, and are not tested on low-level tasks of recalling statistic values. The analysis process was also slower with animated than static traces as participants needed to replay the animation, and replaying is the only way of reinspection of animations despite still resulting in a fleeting reception of information \cite{tversky2002animation}. In addition, between two static displays with trends, trend analysis is more accurate when using a small multiple format that shows individual trends without overlap than when using one single display that combines and overlaps all trends.

While motion helps visualize changes in data, there is a limit to the amount of individual moving objects we can process at a time. When tracking multiple objects across space and time, we can typically track up to four objects. When we attempt to also remember the history of the objects' features (e.g. how the colors, sizes, and trajectories of the data points change), the capacity is even lower \cite{horowitz2007mit}. Thus, such limits in the motion processing capacities restrict the motion information we can process and analyze simultaneously. However, the task people do in this work is not recalling statistic values in visualizations\revision{, judgments such as summary statistics regarding multiple moving objects could still be overwhelming when the dataset is large}.

\subsection{Dynamic Visualizations}

With more visualizations being \revision{displayed on screen than on paper}, more opportunities arise for visualization designers to \revision{use} animation to leverage the power of our visual perceptual system for data storytelling. 
Animated visualizations can deliver a larger amount of information to the viewer compared to their static counterparts within the same number of pixels~\cite{heer2012interactive}.
They can help users more effectively extract summary statistics and increase data engagement \cite{kale2018hypothetical, wang2020cnn, lee2013sketchstory}. 
Animated transitions can help users keep track during data exploration such as pan and zoom operations \cite{bederson1999does}.  \revision{There is often a mismatch in people's preference and performance regarding animated visualizations. People prefer animated visualizations over static ones for being more engaging and more obvious in showing changes, and they tend to be more confident when learning through animated visualizations \cite{boyandin2012qual, brehmer2020trend, robertson2008effectiveness}. However, when designed carelessly, animated visualizations can overwhelm viewers and impair insights compared to static displays (e.g., \cite{hegarty2003roles, robertson2008effectiveness}). It remains unexplored how different designs of animated visualizations would affect analytic tasks.}   

Researchers have begun to explore animation perception in visualizations. Chalbi et.al \cite{chalbi2019common} tested dynamic and stationary variables (position, size, luminance) in conflict conditions and observed how participants group objects. They confirm that the Law of Common Fate can be expanded to dynamic variables with motion being strongest, and their relative weight varies in different real-world settings.
Chevalier et.al. \cite{chevalier2014not} tested the effect of staggering (delay) in animated transition in different complexity by asking participants to track a set of dots. It shows staggering only works in limited conditions.
Robertson et.al. \cite{robertson2008effectiveness} found that trace visualization (like Gapminder) is good at helping people detect abnormal trends.
There is also work on how people understand or design animated visualizations. 
Thompson et al. \cite{thompson_understanding_2020} proposed a taxonomy for ``Animated Data Graphics'' in three dimensions: object, graphic, and data, and divided common transitions into these dimensions. 
They studied how participants break down the animation into dimensions, their preference for transition types, and how they author animations themselves.
Thompson et.al \cite{thompson2021data} proposed an interactive GUI, Data Animator, for animated data creation, leveraging transition from one data view to another. It allows edition on staging and staggering, as well as visual aids on object pairing in different data views. A usability study was conducted to test the completion time of different tasks. Zong et al. \cite{zong2022animated} also created a visualization program language that enhances vega-lite to support animation, and compared the user experience with existing techniques like Data Animator \cite{thompson2021data}.

\revision{While excessive motion information in animated visualizations would likely overwhelm viewers, researchers see the benefits of animations for being enjoyable and depicting changes. With better design, especially targeting specific perceptual tasks and dataset sizes, there is great potential for animations to maintain their benefits and more effectively deliver their underlying messages.}

\begin{figure}[!t]
    \centering
        \includegraphics[width=0.9\columnwidth]{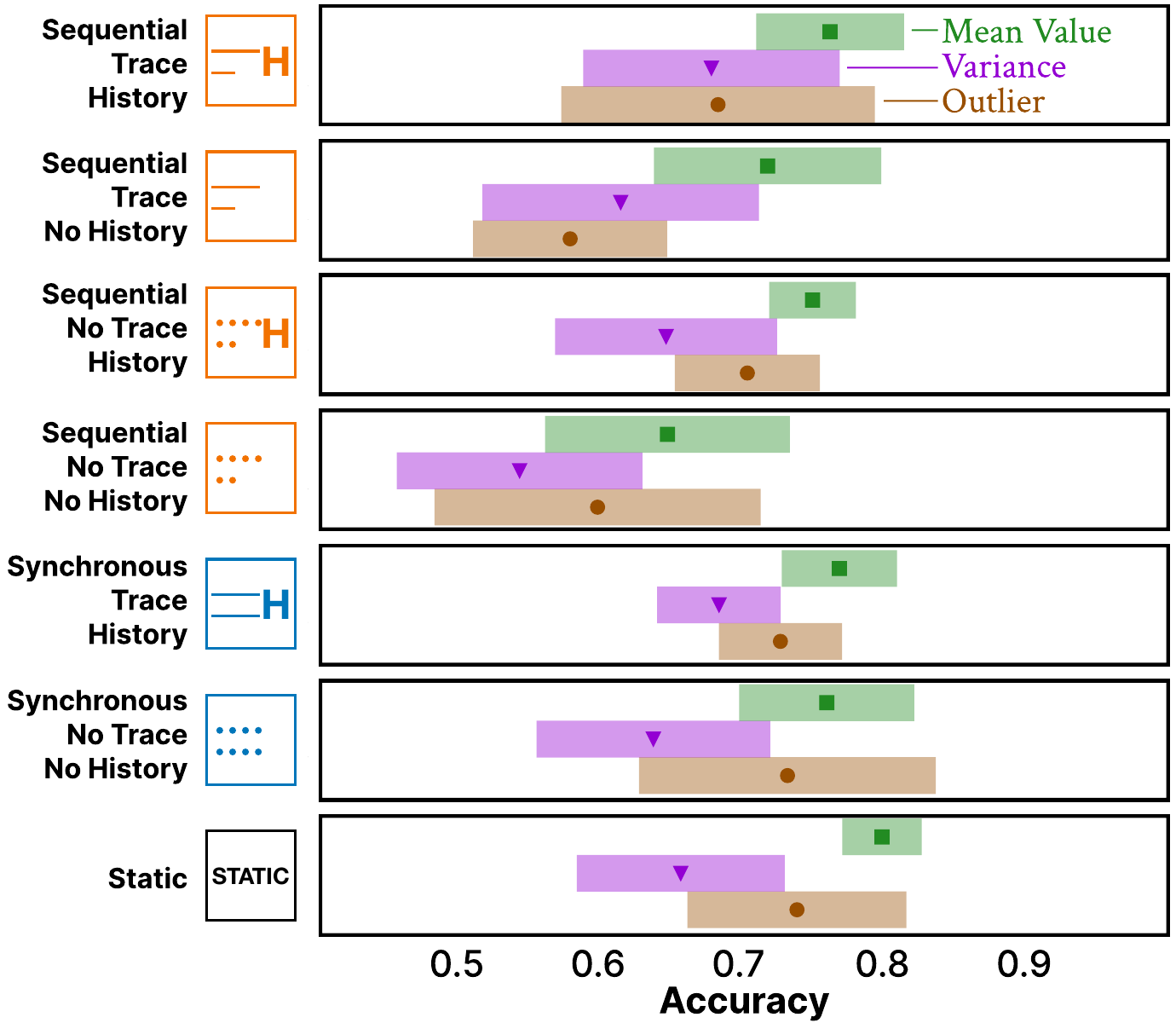}
        \caption{\revision{Recalling mean value has the highest accuracy in all conditions.}}
        \label{fig:figure3a-overall}
        \vspace{-2em}
\end{figure}

\begin{figure}[!t]
    \centering
     \includegraphics[width=0.9\columnwidth]{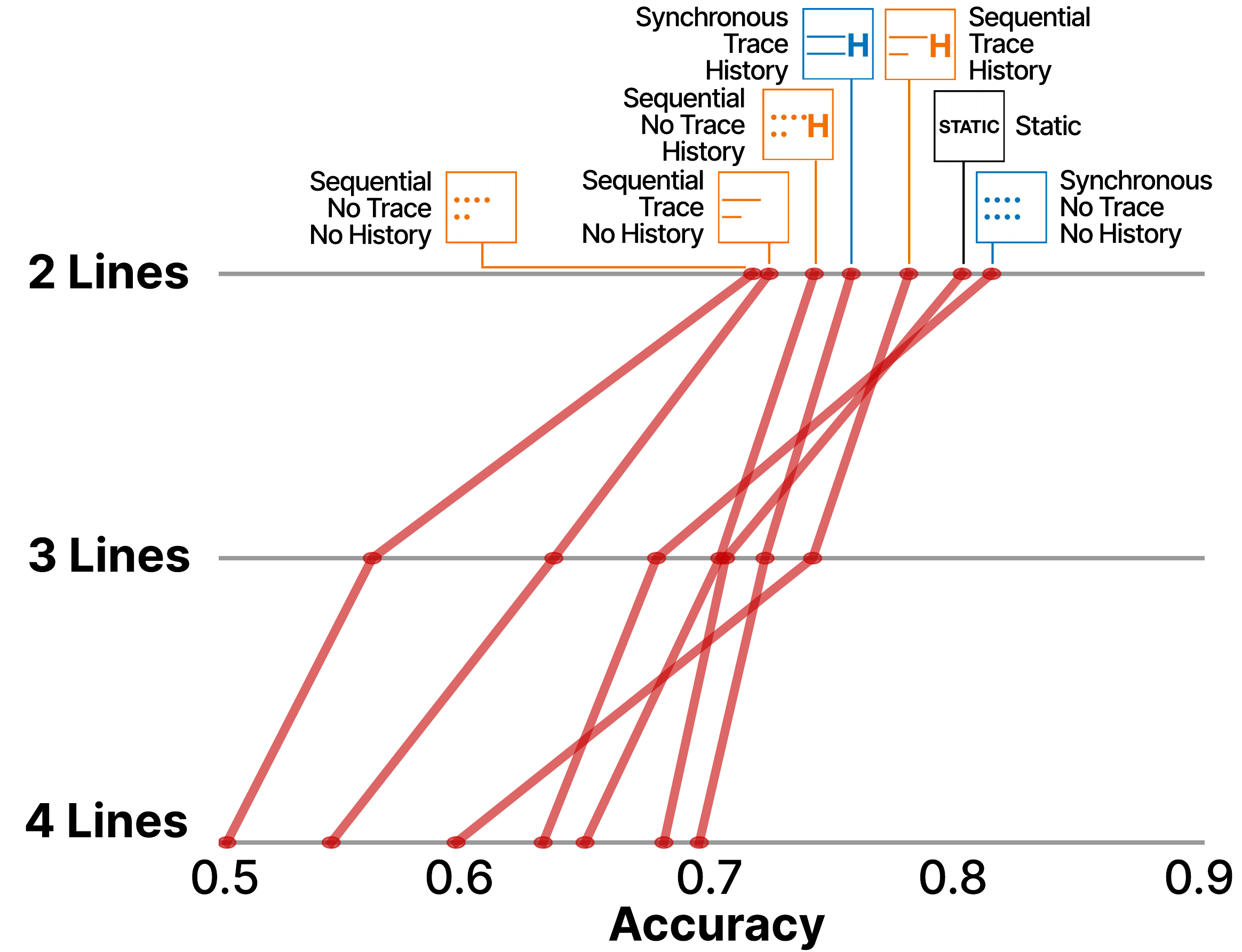}
    \caption{\revision{Accuracy drops with more lines in all conditions.}}
    \label{fig:figure3b-lines}
    \vspace{-1em}
\end{figure}

\begin{figure}
    \centering
    \setlength{\abovecaptionskip}{0pt}
    \includegraphics[width=\linewidth]{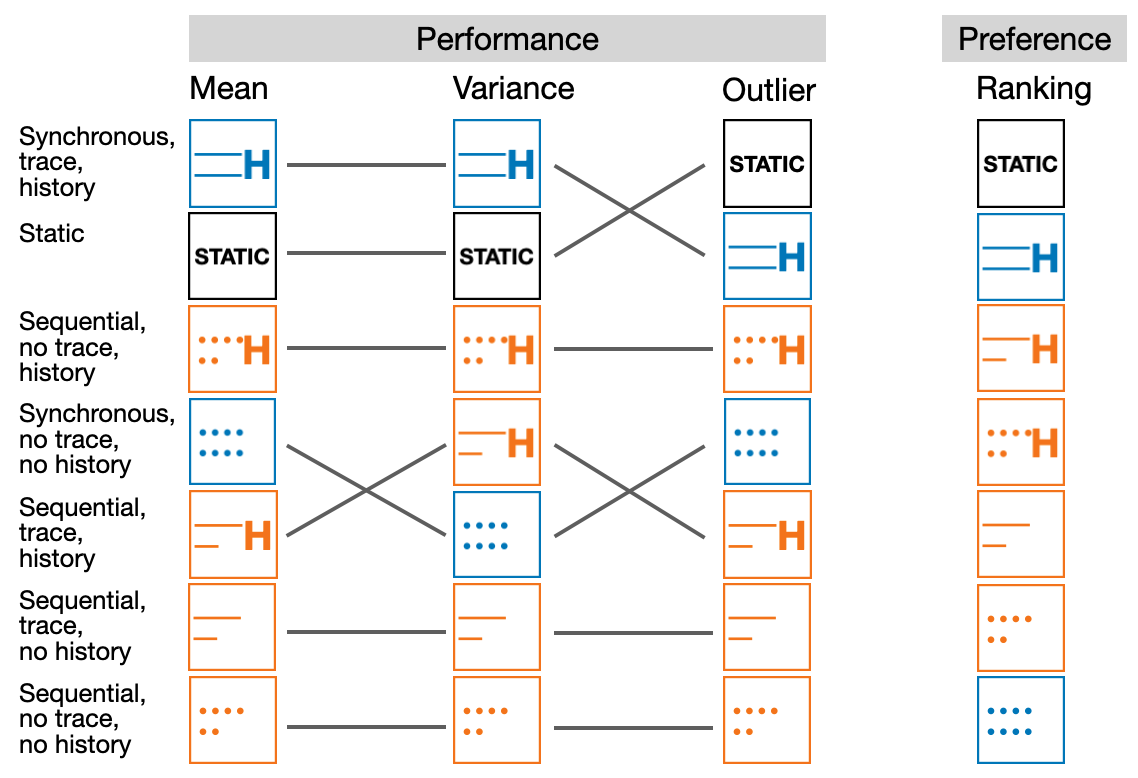}
    \caption{Performance and preference rankings of 4-line dynamic series across conditions. 
    The conditions on top mean higher performance or stronger preference.
    We observed asymmetry between subjective preference and objective performance.}
    \label{fig:performancePreferenceRank}
    \vspace{-2em}
\end{figure}

\section{Exp 1 Amount of Data and Style Preferences}
\label{Exp1}

We first examined how different animation types might benefit or hurt performance on perceptual tasks depending on the amount of data for the readers to process. We also measured their subjective preference for animation styles. 
We designed visualizations animating two, three, or four lines, across combinations of animation techniques. We identified three predominant animation techniques from perception and visualization research to investigate: staging, tracing, and history \revision {preservation} ~\cite{robertson2008effectiveness} \cite{fisher2010anim}.
We manipulated staging by either showing the data values in the visualization sequentially or synchronously, manipulated tracing by either displaying data values' movement trajectory or not, and manipulated history \revision {preservation} by turning the final outcome of what the data looked like after the animation plays through on or off. 

As shown in Figure~\ref{fig:teaser}, we generated six animation conditions: 
\icon{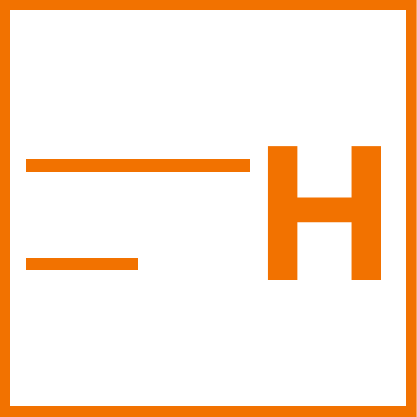}\textit{Sequential, Trace, History}; 
\icon{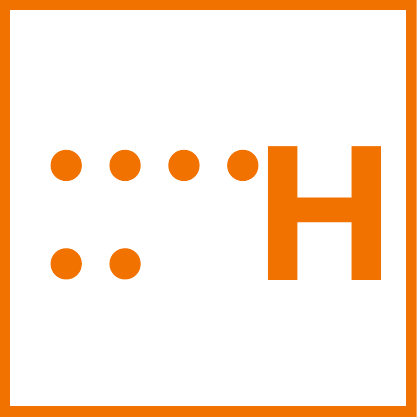}\textit{Sequential, No trace, History}; 
\icon{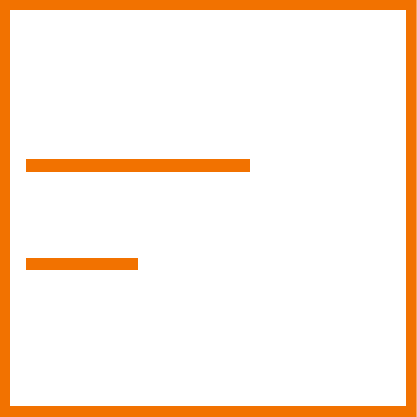}\textit{Sequential, Trace, No History};
\icon{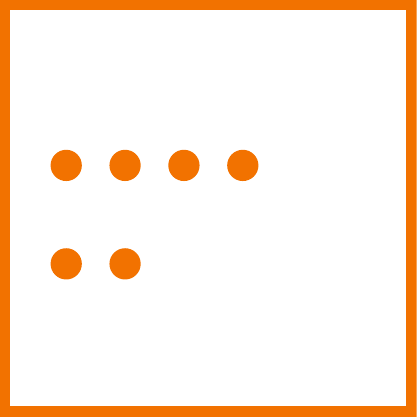}\textit{Sequential, No trace, No History};
\icon{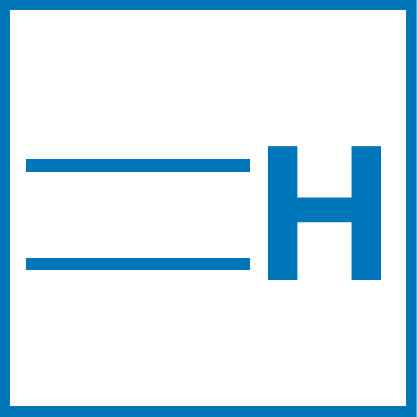}\textit{Synchronous, Trace, History}; and
\icon{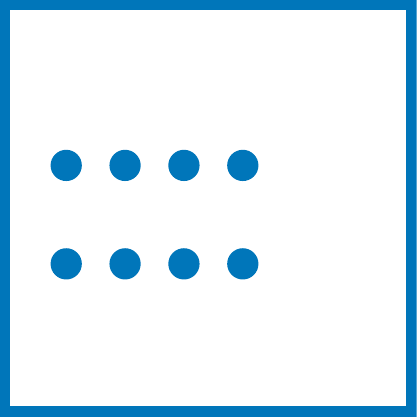}\textit{Synchronous, No trace, No History}. 
Two manipulations from a complete permutation of the three techniques, \textit{Synchronous, no trace, history} and \textit{Synchronous, trace, no history}, were not possible to execute (showing history conflicts with tracing in synchronous lines), and thus were not created as conditions in this study. 
We instead included a static visualization showing all the data at once as a control, resulting in \textbf{seven} conditions total.
\vspace{1mm}

\noindent \textbf{Participants and Safety Measures:} We recruited 189 participants who are fluent in English from Prolific.co~\cite{palan2018prolific}.
The study took on average 40 minutes and participants were compensated at the rate of \$12 per hour. 
\revision{Based on existing works~\cite{borgo2017crowdsourcing},~\cite{borgo2018information}, we implemented two safety measures for experimental quality control. First, we filtered out participants who spent less than 550 seconds, based on the average completion time of 1226 seconds and a standard deviation of 555.1 seconds (people who spent less than 1.5 standard deviations away from the mean time were excluded). Second, participants who provided abnormal responses (people who selected the same answer for all trials or provided nonsensical data in their free response) are excluded. }

\subsection{Experiment Design} 
\label{2design}

We synthesized data to generate time series line charts with two, three, or four lines. 
To ensure generalizability, our synthetic dataset covers a range of data distributions and relations between the means, variances, and the number of outliers.
The synthesized means, variances, and outliers (most significant outliers) were designed to be uniformly sampled from two pools: \textbf{high pool} or \textit{low pool}, based on summary statistics generated from the real-world dataset for external validity (see Section \ref{Exp4}).
For all charts, the x-axis ranged from 0 to 100, and the y-axis from 100 to 700. 
We calculated the mean values and standard errors (to represent variance) for each subset of data.
The means in \textit{high pool} are filtered to be larger than the median of all mean values, and those in \textit{low pool} are filtered to be smaller than the median.
Similarly, the variance values (represented with standard deviation) are set to be higher than the median of all standard deviations in the \textit{high pool}, and lower than the median in the \textit{low pool}. 
For the most significant outlier, \textit{high pool} means having an outlier while \textit{low pool} means not. If a line has an outlier, its y-axis value would be five times the standard error of the line's data, and its x-value is uniformly sampled from 0 to 100.

We permutated through the high and low pools for these three metrics to generate displays.
The permutation for 6 visualization designs plus the static condition yielded $1512$ conditions (see supplementary materials for details). 
For each display we created 5 color variants to control the effect of color, resulting in $1512\times 5 = 7560$ animated visualizations. 
\revision{We used a between-subject design to examine the effect of animation conditions and line numbers on task accuracy.} 
We controlled for animation display time and kept it consistent across all conditions to be 5 seconds (with 1 second of blank frame).
We further tease apart the effect of animation speed in Experiment 2 and 3.

\subsection{Experiment Procedure} 
\label{sec: exp1-procedure}
After consenting to the study, to ensure that everyone sees the stimuli in the same resolution and size, participants adjusted the size of a visualization by resizing a rectangle to fit the size of a credit card on the web page, according to virtual-chinrest by Li et al. \cite{li2020controlling}. 
Participants then went through a short training on how to compare means, variances, and the most significant outliers in time series. 
Next, they went through a practice trial where they viewed an animated time series visualization (with 4 lines appearing sequentially and trace and history both turned 'on'), and identified the line with the highest mean, variance, and outlier.
Participants can study the visualization for unlimited time before proceeding to the experiment. 

For the main experiment, participants viewed an animated display for five seconds and, on a separate page, completed three perceptual tasks involving comparing the means, variances, and the most significant outliers (operationalized as identifying the line with the most 'significant spike' ) of the lines.
Specifically, they were told to identify the line with the highest mean, variance, and the most significant outlier value. 
They repeated this task 40 times, with each display randomly sampled from the 7560 visualizations described in Section \ref{2design}. 

We also asked the participants to rank the 7 visualization conditions (6 animations and 1 static) by arranging them in a horizontal box from left to right, with the left-most display being the most disliked and the right-most display being the most preferred. 

The experiment finished with demographic questions asking for participants' level of education and experiences coming across and designing animated charts.

\subsection{Results: Amount of Data and Animation Overview}

We first took an overview to examine the effect of the amount of data on performance across three tasks, and used a logistic regression model to predict performance with the number of lines crossed with three tasks.
As shown in Figure~\ref{fig:figure3b-lines}, we found a main effect of the number of lines, such that participants were the most accurate when completing perceptual tasks with two lines,  1.66 times more accurate than tasks with three lines, and 2.32 times more accurate than tasks with four lines.
These perceptual tasks were the most difficult when there were four lines in the display, where participants' performance was significantly worse compared to displays with three lines ($OR=1.40$, $p<0.0001$). 
Overall, participants performed the most accurately on the outlier task, 1.23 times more accurately than the mean comparison task and 1.62 times more so than the variance comparison task ($p$<$0.001$).
They performed the worst on the variance comparison task, and were only 1.32 times as likely to correctly respond as compared to the mean comparison task ($p$<$0.001$), see Figure~\ref{fig:figure3a-overall}.

\subsection{Results: Effect of Animation Conditions}

We first took an overview of a logistic model predicting performance for each task based on whether staging, tracing, and history preservation are manipulated. 
Overall, we found that staging hindered task performance (OR=$0.696$).
Tracing also hindered task performance, although the effect seemed small (OR=$0.978$).
Showing history improved task performance by 1.36 folds. 

Next, we more systematically examined performance on the three perceptual tasks across the six animation conditions and three line numbers. 
This allowed us to identify the most effective animation technique for each task.
We report the most significant takeaways from our analysis. 
Refer to supplementary materials for detailed statistics. 
\vspace{1mm}

\noindent \textbf{Mean Task:}
We show the participants' performance in comparing the mean value in Figure~\ref{fig:figure3a-overall}. 
The best-performing condition is \icon{sync_notra_nohis} \textit{synchronous, no trace, no history} for 2 lines, \icon{seq_tra_nohis}\textit{sequential, trace, history} for 3 lines, and \icon{sync_tra_his}\textit{synchronous, trace, history} for 4 lines. 
Generally, synchronous animations and static visualization perform well in recalling the highest mean value. 
For sequential animations, keeping histories can improve people's performance, especially when there are more lines in the animation.
\vspace{1mm}

\noindent \textbf{Variance Task:}
As shown in Figure~\ref{fig:figure3a-overall}. The best-performing condition is \icon{seq_notra_his}\textit{sequential, no trace, history} for 2 lines, \icon{seq_tra_his}\textit{sequential, trace, history} for 3 lines, and both \icon{sync_tra_his}\textit{synchronous, trace, history} and \icon{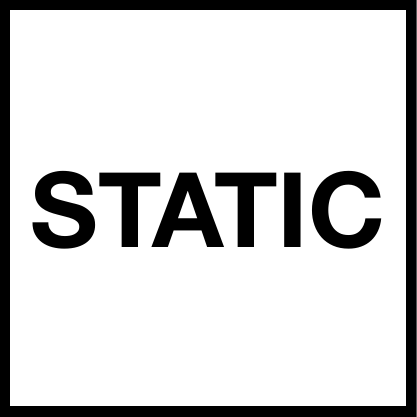}\textit{static} for 4 lines. 
Generally, static visualization and animations with traces perform well in comparing the variance.
For both sequential and synchronous animations, keeping traces can improve people's performance. For sequential animations, keeping histories can also improve people's performance.
\vspace{1mm}

\noindent \textbf{Outlier Task:}
We show the participants' performance of recalling the most significant outlier in Figure~\ref{fig:figure3a-overall}. The best-performing condition is \icon{sync_notra_nohis}\textit{synchronous, no trace, no history} for 2 lines, \icon{seq_notra_his}\textit{sequential, no trace, history} for 3 lines, and \icon{static}\textit{static} for 4 lines. Static and animations with histories have generally high performance in the task of recalling the most significant outlier. For both sequential and synchronous animations, keeping histories can improve people's performance, but keeping both trace and history in sequential animations can lower people's performance when there are more than 3 lines.
\vspace{1mm}

\noindent \textbf{Subjective Preference:}
We conducted a one-way Kruskal-Wallis ANOVA with a Dunn's post-hoc analysis to compare preference for each animation condition (p-values are corrected via Bonferroni's method to account for multiple comparisons).
Overall, participants preferred the \icon{static} \textit{Static} display the most ($Mean_{rank}$=$2.80$, $SE$=$0.135$), followed by \icon{sync_tra_his} \textit{Synchronous, trace, history} ($Mean_{rank}$=$3.04$, $SE$=$0.129$, $p$=$0.00975$). 
\icon{sync_notra_nohis} \textit{Synchronous, no trace, no history} was the least preferred ($Mean_{rank}$=$5.30$, $SE$=$0.129$) and \icon{seq_notra_nohis} \textit{Sequential, no trace, no history} ($Mean_{rank}$=$5.30$, $SE$=$0.143$, $p$=$0.979$).
\vspace{1mm}

\noindent \textbf{Comparing Subjective and Objective Measures:}
As shown in Figure~\ref{fig:performancePreferenceRank}, participants varied in how well they completed the mean, variance, and outlier tasks depending on the animation condition. 
Their reported animation preferences also did not match their objective performance. That is, participants did not always complete the perceptual tasks with the highest accuracy under the more preferred animation conditions. 
We performed a Kendall's Tau correlation test for performance rankings of the three tasks without considering the number of lines. The correlation between the performance rankings of mean and variance was 0.33, the correlation between the performance rankings of mean and significant outlier was 0.52, and the correlation between the performance rankings of variance and significant outlier was 0.05. 

We report participants' rankings of the visualizations in Figure~\ref{fig:performancePreferenceRank}.
\icon{static}\textit{Static} was most preferred, followed by \icon{sync_tra_his}\textit{synchronous, trace, history}. These were also the top two visualization conditions for the accuracy performance of the three tasks. \icon{seq_notra_nohis}\textit{Sequential, no trace, no history} and \icon{sync_notra_nohis}\textit{synchronous, no trace, no history} were the least preferred.
\vspace{1mm}

\noindent \textbf{Discussion:} 
The conditions \icon{sync_notra_nohis}\textit{synchronous, no trace, no history} were not preferred by most participants, but people are performing well in the perceptual tasks, especially when comparing the mean and the most significant outliers. 
Because this condition yielded the highest performance in mean and outlier tasks, we suspect animation styles to have their own affordances for different visual tasks, similar to other encoding channels. 
This motivates us to investigate what analytic task each speed and animation technique affords in subsequent studies. 

\begin{figure*}[!t]
\vspace{-2em}
    \centering
     \setlength{\abovecaptionskip}{0pt}
    \includegraphics[width=\textwidth]{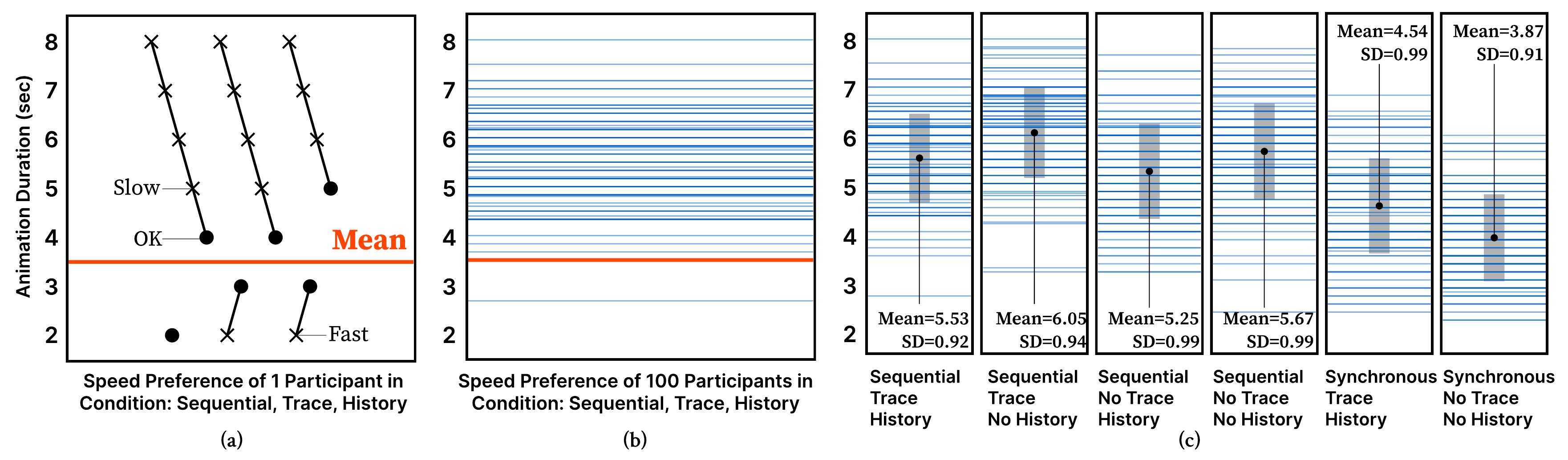}
    \caption{Experiment 2,
    Figure (a) shows one participant's response for condition \icon{seq_tra_his} \textit{sequential, trace, history}. The mean value of speeds where “OK” is reported is taken as the preferred speed for that participant. 
    Figure (b) shows the distribution of the preferred speed across 100 participants, where the speed derived in Figure (a) is marked as orange. 
    Figure (c) shows the distribution of preferred speed for different designs in each column with strip plots like figure (b). It also shows the mean and standard deviation of the preferred speed distribution in each design.}
    \label{fig:speedPreference}
    \vspace{-1em}
\end{figure*}

\begin{figure*}[!t]
    
    \centering
    \setlength{\abovecaptionskip}{0pt}
    \includegraphics[width=\textwidth]{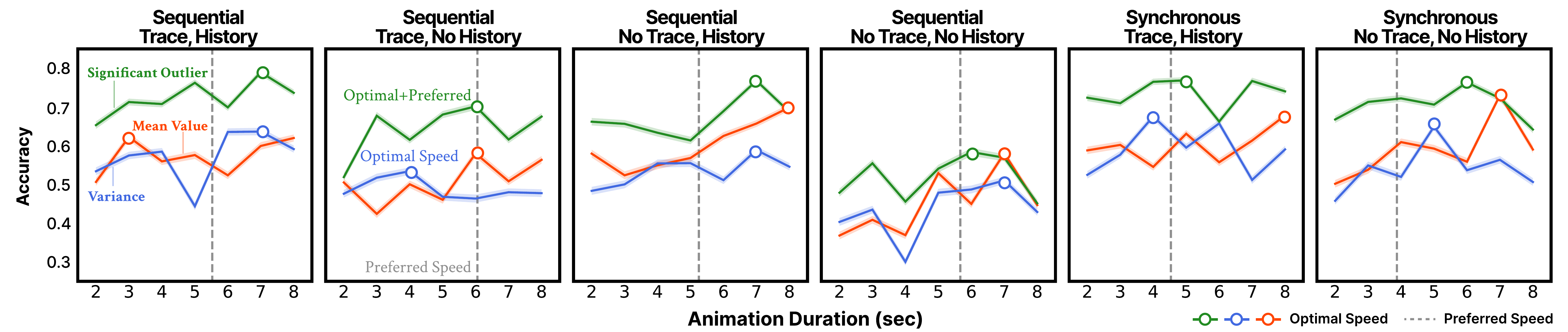}
    \caption{Experiment 2, participants’ accuracy of different visualization conditions in mean, variance, significant outlier tasks. The animation display time preferred by participants is marked with stars, and the animation display time with optimal performance is marked with dots.}
    \label{fig:speedAccuracy}
    \vspace{-2em}
\end{figure*}

\section{Exp 2 Preference \& Performance Under Speeds}
\label{expSpeedPref}

To design animated displays with speed as an encoding channel, we first identified which speeds are the most subjectively preferred by visualization readers.
We then examined readers' performance for the mean, variance, and outlier tasks under these speeds. 

\subsection{Participants, Experiment Design, and Procedure} 
We recruited 100 participants from Prolific.co~\cite{palan2018prolific}, with the same filtering criteria and \revision{safety measures for quality control} as those in Experiment 1.
The study took about 38 minutes.
Participants were compensated at the rate of \$12 per hour. 

\revision{We conducted a within-subject experiment to study the effect of animation conditions and speed on accuracy and preference.} 
We created dynamic time series visualizations showing the movement of four lines, with line colors randomly assigned to each line. 
We manipulated the animation speed with display times ranging from two to eight seconds with one-second increments (seven unique display times in total). 

In the first phase, we adopted the psychophysics method of limits~\cite{elliott2020design}. 
Participants viewed a dynamic visualization playing for either two seconds (maximum speed) or eight seconds (minimum speed). 
They indicated whether the animation was too slow, too fast, or `OK'. 
Depending on their response, we increased or decreased the animation speed until they selected `OK'.
We repeated this task over six trials, randomly dispersed across all the other trials, alternating between starting from slow or fast for all six animation conditions.  

In the second phase, we showed the participants displays in the six animation conditions, across all intervals of display times ($6 \times 7 = 42$ displays).  
For each display, participants completed the mean, variance, and outlier tasks.

\subsection{Results: Subjective Preference}

We first computed each person's preferred speed over the six trials they viewed for each animation condition.
Figure~\ref{fig:speedPreference}(a) shows one participant's data for a single animation condition \icon{seq_tra_his}\textit{sequential, trace, history}. In this example, for the first trial, they saw a 2-second display and indicated that the animation speed was `OK'. 
They then saw an 8-second display and indicated the animation was `too slow' until the speed was lowered to 4 seconds. 
Next, they saw a 2-second display and indicated the animation was `too fast' until the speed was increased to 3 seconds, and so on.
Speeds marked as `OK' are averaged to generate a mean preferred speed for this animation and participant. 
We repeated this process for all participants (see Figure~\ref{fig:speedPreference}(b)) and all animation conditions (see Figure~\ref{fig:speedPreference}(c)). 

We first took an overview of the effect of staging, tracing, and history preservation techniques on preference via an ANOVA predicting preferred play speed with these three variables (two levels for each: on/off), with post-hoc pair-wise comparisons.

We consider synchronous displays as not staged and sequential displays as staged. 
We found participants overall preferred a shorter display time for 
synchronous displays ($Mean=4.20$, $SE=0.064$) compared to 
sequential displays  ($Mean=5.60$, $SE=0.046$, $p_{adj}<0.0001$). 
Sequential displays stage the line animations, therefore it takes four times as long for a sequential display to play out compared to a synchronous display showing the same four lines, assuming the lines move at the same speed.
If participants have an absolute preference for line movement speeds, we would expect their preferred display time for the sequential display to be four times that of the synchronous display. 
However, our results indicated that to not be the case. 
On average, participants preferred the lines in a sequential display to move three times faster than the lines in a synchronous display. 

For tracing, participants preferred longer animation display time for 
displays with trace ($Mean =  5.16$, $SE = 0.056$) compared to 
displays without trace ($Mean =  4.64$, $SE = 0.055$, $p_{adj} <0.0001$). 
This suggests that participants preferred the displays with trace to play slower than displays without trace.

However, for history \revision {preservation}, participants preferred shorter animation display time for 
displays with history ($Mean =  4.78$, $SE = 0.056$) compared to 
displays without history ($Mean =  5.03$, $SE = 0.056$, $p_{adj} <0.0001$). 
This suggests that participants preferred animation with history to play faster.

For clarity of result interpretation, we also conducted a one-way ANOVA comparing the six conditions to better understand participants' relative speed preference across these conditions. 
Overall, participants preferred significantly different speeds across the six animation styles ($F=111.10$, $p$<$0.0001$). 
We conducted a post-hoc analysis with adjustments for multiple pair-wise comparisons. We report significant differences here, and detailed statistics can be found in the supplementary materials.
Our data suggests that participants preferred the longest display time ($t$ = 6.05s) for the \icon{seq_tra_nohis}\textit{sequential, trace, no history} condition. 
The next tier of display time includes the \icon{seq_notra_nohis}\textit{sequential, no trace, no history} ($t$ = 5.67s) and the \icon{seq_tra_his}\textit{sequential, trace, history} ($t$ = 5.53s) conditions.
This is followed by the \icon{seq_notra_his}\textit{sequential, no trace, no his} condition, where participants preferred a moderate display time ($t$ = 5.25s) compared to all other conditions.
They preferred the \icon{sync_tra_his}\textit{synchronous, trace, history} to play at the second shortest time of $4.54$s.
They preferred the display time to be the shortest for the \icon{sync_notra_nohis}\textit{synchronous, no trace, no history} condition, at $3.87$s.
\vspace{1mm}

\noindent \textbf{Discussion:} Animation design has a significant effect on people's subjective preference for animation display speed. Generally, participants preferred the lines in a synchronous display to move three times slower than the lines in a sequential display. Regarding trace and history, participants preferred the animations with trace to play slower than those without trace, but preferred animations with history to play faster than those without history. It suggests that even though trace and history both involves leaving the lines on the screen for a certain amount of time, people have opposite preference for them and they should be treated separately in design. When the trace is displayed, participants have more information to look at but are limited to animation duration, so they might need more time. 
On the other hand, preserving the history until the end of the entire animation might make participants feel more ``assured'' that they will have history information later for reference, so there is no need for longer display time. 
\revision{This suggests that more complex perceptual heuristic mechanisms might be involved.
For example, participants might be sensitive to both the horizontal speed of the line movement, and the speed of jumps between data points, as drastic changes capture their attention.}

\subsection{Results: Speed and Task Performance}

Next, we examined participants' objective performance across all seven levels of animation speeds (2 to 8 seconds).
We conducted a logistic regression predicting task accuracy with the type of perceptual task (mean, variance, significant outliers), animation conditions (with or without staging, tracing, history), and animation display time.

Overall, participants performed most accurately on the outlier task, 1.62 times more so than the mean comparison task and 1.82 times more so than the variance comparison task ($p$<$0.001$).
They performed worst on the variance comparison task, being 0.89 times as likely to respond correctly as on the mean comparison task ($p$<$0.001$).

Animation speed has a main effect ($p$<$0.0001$) that with each second of increase in animation display time, the accuracy increases by 1.05 folds across all perceptual tasks. 
In terms of staging, tracing, and history preservation, 
staging the animation (displaying lines sequentially) \textit{decreases} the accuracy by 1.34 folds, while tracing \textbf{increases} by 1.08 folds and history \textbf{increases} by 1.38 folds (all $p$<$0.001$).

Next, we systematically examined perceptual task performance across the six animation conditions for the three tasks.
This allows us to identify the most effective animation technique for each task.
Again, we report the most significant takeaways from our analysis. Details can be found in the supplementary materials. 
\vspace{1mm}

\noindent \textbf{Mean Task.} 
For the mean task, we see a small effect of time, such that increasing display time by one second would increase the odds of getting the correct answer by 1.07 fold ($p$<$0.001$). 
Overall, the animation conditions \icon{sync_tra_his} \textit{synchronous, trace, history} leads to the highest performance on the mean task.
The animation condition \icon{seq_notra_nohis} \textit{sequential, no trace, no his} resulted in the lowest performance.
\vspace{1mm}

\noindent \textbf{Variance Task.}
We see a small effect of time, such that increasing display time by one second would increase the odds of getting the correct answer by 1.03 fold ($p$<$0.001$). 
Overall, similar to the mean comparison task, the animation condition \icon{sync_tra_his} \textit{synchronous, trace, history} leads to the highest performance, and \icon{seq_notra_nohis} \textit{sequential, no trace, no his} leads to the lowest performance.

\vspace{1mm}

\noindent \textbf{Outlier Task.}
For the outlier task of identifying the line with the most significant outlier, we observed a small effect of time, such that increasing display time by one second would increase the odds of getting the correct answer by 1.04 fold ($p$<$0.001$). 
Overall, the animation condition \icon{sync_tra_his} \textit{synchronous, trace, history} leads to the highest performance on the mean task, and the condition \icon{seq_notra_nohis} \textit{sequential, no trace, no history} leads to the lowest performance.

\noindent \textbf{Discussion:} 
\icon{sync_tra_his}\textit{Synchronous, trace, history} had the highest performance for all three tasks 
while \icon{seq_notra_nohis}\textit{sequential, no trace, no history} had the lowest performance for all tasks. 
Increasing display time only had a small effect on increasing performance 
and cannot compensate for the low performance brought by other conditions. 

\subsection{Results: Comparing Preference and Performance}

Overall, across the six animation conditions and the three perceptual tasks, we observed a discrepancy between participants' subjective display time preference compared to the display time that affords the highest accuracy in the perceptual task.
\vspace{1mm}

\noindent \textbf{Mean Task:} Participants performed the mean comparison tasks with the highest accuracy (65.0\%, $SE$=$0.0024$) when the display time is 6.50s ($SE$=$0.070$), which is significantly different from their preferred display time of 5.15s ($SE$=$0.049$, $t$=$-15.788$, $p$<$0.001$). 
Their accuracy at their preferred display time is only 55.4\% ($SE$=$0.0026$).
\vspace{1mm}

\noindent \textbf{Variance Task:} Participants performed the variance comparison tasks with the highest accuracy (60.0\%, $SE$=$0.0025$) when the display time is 6.50s ($SE$=$0.056$), which is significantly different from their preferred display time of 5.15s ($SE$=$0.049$, $t$=$-6.90$, $p$<$0.001$). 
Their accuracy at their preferred display time is only 51.6\% ($SE$=$0.0024$). 
\vspace{1mm}

\noindent \textbf{Outlier Task:} Participants performed the variance comparison tasks with the highest accuracy (73.2\%, $SE$=$0.0029$) when the mean display time is 6.50s ($SE$=$0.070$), which is significantly different from their preferred display time of 5.15s ($SE$=$0.049$, $t$=$-15.79$, $p$<$0.001$). 
Their accuracy at their preferred display time is only 66.2\% ($SE$=$0.0033$).
\vspace{1mm}

\noindent \textbf{Discussion:} Across all three tasks, participants did not perform the best at their preferred display time, suggesting a discrepancy between subjective v.s. objective measures of optimal animation speed.

\begin{figure}[!t]
    \centering
    \setlength{\abovecaptionskip}{0pt}
    \includegraphics[width=\linewidth]{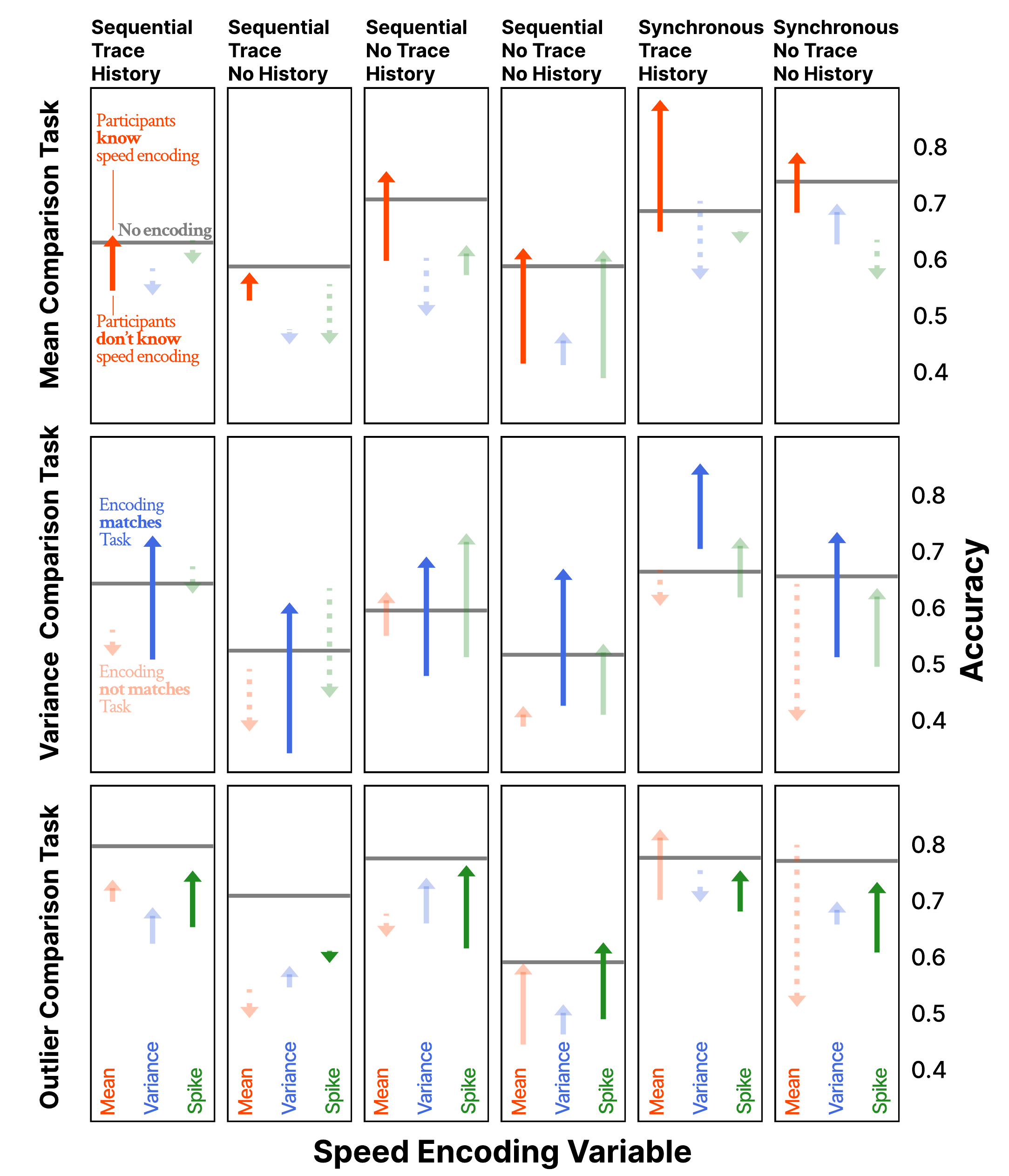}
    \caption{Participants’ accuracy with different animation design, task, and encoding in Experiment 3. 
    For each arrow, the beginning value represents the accuracy when participants were \textit{not} explicitly told about the meaning of the speed encoding, and the ending value represents accuracy when they were explicitly told. 
    If the arrow points upward, it suggests that being explicitly told about the meaning of the speed encoding improved their task performance, while pointing downwards suggests knowing the speed encoding hurts their performance.}
    \label{fig:exp3}
    \vspace{-2em}
\end{figure}

\section{Exp 3 Speed as a Visual Encoding Channel}
\label{exp3b}

With the optimal speeds identified, we now closely examine how motion-based data encoding channels can benefit or stymie perceptual task performance, across the three tasks and six animation conditions. 
We explore the potential benefits or hindrances to perceptual task performance by redundantly encoding animation speed with y-axis position, data variance, and the most significant outliers.

\subsection{Participants, Experiment Design, and Procedure} 
The study took about 15 minutes, and participants were compensated at the rate of \$12 per hour.
We recruited 100 participants from Prolific.co~\cite{palan2018prolific}, with the same filtering criteria and \revision{safety measures for quality control} as those in Experiment 1.

We randomly generated dynamic line charts showing four lines following the same algorithm as that in Experiment 1, for all six animation conditions. 
We identified the animation speeds associated with the highest performance on perceptual tasks based on results from Experiment 2. 
For each animation condition, we manipulated the movement speed of each line via three encoding choices to associate speed with one of the three data attributes (y-position, variance, and number of outliers), to generate 6*3 = 18 animated visualizations: 
\vspace{1mm}

\noindent \textbf{Choice \#1} Speed is redundantly encoded with the average y-axis position (lines with higher mean value move faster).

\noindent \textbf{Choice \#2} Speed is redundantly encoded with the variance (lines with higher variance move faster).

\noindent \textbf{Choice \#3} Speed is redundantly encoded with the significant outliers (lines with a more significant outlier move faster).
\vspace{1mm}

\noindent These encoding choices will help us observe whether encoding a specific attribute would increase performance accuracy when people extract statistics from that particular attribute.

\revision{We conducted a within-subject experiment to study the effect of knowing, encoding channels, and animation conditions on accuracy.} 

\vspace{2mm}

\noindent \textbf{Animation Speed:} We determined the specific animation speeds by taking the speeds under which participants in Experiment 2 performed the most accurately. 
Take the versions where speed is redundantly encoded with the most significant outlier as an example.
For the \icon{sync_tra_his}\textit{synchronous, trace, history} condition, we chose the animation that plays for 5 seconds, as it is the display time that elicits the highest accuracy in the outlier task.
For the \icon{seq_tra_his} \textit{sequential, trace, history} condition, we chose the animation that plays for 7 seconds (see Figure~\ref{fig:speedAccuracy}).

However, since we are treating speed as an encoding channel, each line in the display should have a different animation speed corresponding to the data attribute (mean, variance, and significant outlier) 
with which speed is encoded, without altering the overall animation time.  
To achieve this, we scaled the display time for each line by the attribute where the speed is encoded. 
For sequential displays, keeping the overall animation display time means that each line's display time should be scaled.
To determine the scaling interval, we referenced the data collected on participants' animation speed \textit{preference} from Experiment 2.
The preference data shows a standard deviation of one second,
which suggests that a one-second change in overall animation display time is noticeably visible to a viewer. 
Considering that the fastest preferred display time for sequential animation was three seconds, we scaled each line's display time to ensure their interval was 1/3 of the original display time. Take the \icon{seq_tra_his} \textit{sequential, trace, history} condition that each line plays for 7/4=1.75 seconds as an example: the display time of lines with most to least significant outliers would be $1.75\times [0.5,0.5+1/3,0.5+2/3,1.5]$ seconds.
For synchronous displays, adhering to the overall animation display time means that the display time of the slowest moving line should match the overall animation display time.
With this constraint, considering the one-second noticeable difference, and that the fastest preferred display time for synchronous display is five seconds, we scaled each line's display time to ensure the interval is $1/5$ of the original display time. Take the \icon{sync_tra_his} \textit{synchronous, trace, history} condition that all lines play for five seconds as an example: the display time of lines with most to least significant outliers would be $5\times[2/5,3/5,4/5,1]$ seconds.
\vspace{2mm}

\noindent \textbf{Intuitiveness:} To investigate the intuitiveness of the speed-based redundant encoding, we also manipulated whether participants were explicitly told about the meaning of the speed encoding channel. 
To mitigate the effect of the curse of knowledge~\cite{xiong2019curse}, we generated another set of 18 animated displays and showed the two sets to participants in two blocks.
In the first block, participants were not told anything besides to look at the animation, before they completed the mean, variance, and outlier tasks.
In the second block, participants were explicitly told that the display they saw has speed redundantly encoded with either the mean, variance, or significant outliers (i.e., ``higher speed means higher average/variance/spikes'').
Overall, each participant saw 36 animated visualizations (6 conditions x 3 speed encodings x 2 sets).

As with Experiment 1, after consenting to the experiment, participants went through an instruction module training them on how to extract mean, variance, and outlier information from the line displays.
They view the 36 animated visualizations in two blocks, and respond to the mean, variance, and outlier perceptual tasks.

\subsection{Results: Intuitiveness (Effects of Knowing)}
According to Figure~\ref{fig:exp3}, if speed encoding matches the perceptual task, explicitly telling participants the meaning of the speed encoding improves their task performance across all tasks and animation conditions. One exception is that comparing outliers in \icon{seq_tra_nohis} \textit{sequential, trace, no history} has a slight decrease in accuracy. 
If speed encoding does not match the perceptual task, there is no obvious pattern between telling participants the meaning of speed encoding and the change in their performance.

We built a logistic regression model predicting response accuracy for each task, based on whether participants were told explicitly about the meaning of the speed encoding or not. 
Generally, we found speed encoding to be significantly more beneficial when participants were explicitly told its meaning, as illustrated by the higher number of arrows pointing upwards in Figure~\ref{fig:exp3}.
Notably, explicit knowledge of the speed encoding channel varied in how much they benefit each task. 

Explicitly knowing the meaning of speed encoding increased performance the most in the variance comparison task, by 1.31 fold, followed by the mean comparison task (OR=$1.16$), and least so in the significant outlier task (OR=$1.14$, all $p$<$0.001$).
\vspace{1mm}

\noindent \textbf{Takeaway:} 
Overall, using speed encoding will be the most effective when the speed encoding matches the perceptual task, and when you explicitly tell participants what speed represents.

\subsection{Results: Benefits of Speed Encoding}
Next, we examined the overall benefits of adding speed as an encoding channel.
We compared participants' overall performance in this experiment to participants' performance in Experiment 2 when speed was not used as an encoding channel. 
The horizontal black lines in Figure~\ref{fig:exp3} represent the baseline accuracy from Experiment 2. 

We are also interested in understanding whether speed encoding can hurt perceptual task performance through interference.
For example, encoding the average variance with speed could make a mean comparison task more difficult to complete, as participants get distracted by the salient speed encoding that represents variance and become less attentive to mean values. 
For each perceptual task, we constructed a logistic regression predicting the effect of encoding the mean, variance, or significant outliers with speed on task accuracy. 
\vspace{1mm}

\noindent \textbf{Mean Task:} We compared participants' performance on the mean comparison task when speed is encoded with mean values (congruent), variance (incongruent), and the significant outlier task (incongruent). 

Overall, while encoding speed with mean values enhances performance on the mean comparison task (OR=$1.10$), encoding speed with data attributes incongruent with the mean task significantly hurts task performance ($p$<$0.001$).
Namely, encoding speed with variance values \textit{decreases} accuracy on the mean task by 1.25 folds, and encoding speed with the significant outlier attribute \textit{decreases} accuracy on the mean task by 1.13 folds. 
As shown in the top row of Figure~\ref{fig:exp3}, the average positions of the orange arrows, representing speed encoded with mean values, are higher than the black baseline.
The average positions of the blue and green arrows, representing speed encoded with variance and outlier, are lower than the black baseline.
\vspace{1mm}

\noindent \textbf{Variance Task:} Encoding variance values with speed (congruent) improved performance on the variance comparison task by 1.12 folds, and encoding mean values with speed (incongruent) hurt performance on the variance comparison task by 1.20 folds ($p$<$0.001$).
Interestingly, for this task, encoding significant outlier values with speed did not interfere with performance on the variance task despite the incongruence, increasing performance accuracy by 1.07 folds ($p$=$0.002$).
\vspace{1mm}

\noindent \textbf{Outlier Task:} Performance in the outlier task was hurt by all three choices of speed encodings, congruent or not ($p$<$0.001$).
Encoding mean values with speed decreased outlier task accuracy by 1.40 folds, and encoding variance values with speed decreased it by 1.45 folds. 
Surprisingly, even encoding the most significant outlier with speed decreased outlier task accuracy by 1.32 folds.
\vspace{1mm}

\noindent \textbf{Takeaway:} 
Speed encoding helps in the task of recalling mean value and variance, but only when the encoding is congruent with the goal at hand. 
Encoding data attributes in a way that's incongruent 
to the perceptual task generally hurts performance.
Speed encoding is not helpful for significant outlier tasks and should be avoided if the goal is to identify outliers.
\revision{The asymmetrical benefits of speed encoding to analytic tasks also suggests a complex underlying mechanism for visualization animation perception that warrants future investigation.}

\subsection{Results: Staging, Tracing, and History Preservation}

Next, we examine how speed encoding might interact with animation conditions to even augment or diminish performance in perceptual tasks. 
We first took an overview of a logistic model predicting performance for each task based on whether staging, tracing, and history are manipulated.
Overall, we found that when speed encoding was present, staging hindered task performance (OR=$0.74$).
Tracing also hindered task performance, although the effect seemed small (OR=$0.74$).
History improved task performance by 1.49 folds. 
\vspace{1mm}

\noindent \textbf{Discussion:} In Experiment 2, where speed encoding was not used in displays, we found that staging hurt performance while tracing and history helped. These results hold when speed encoding was used in displays, except for the diminished benefits of tracing. 

\subsection{Results: Speed encoding by Animation Condition}

Next, we took a closer look at the relative performance across the six animation conditions to generate task-specific recommendations on which animation style to use to optimize performance. 
We constructed logistic linear regression models for each task, to compare performance under each of the six animation conditions. 
Instead of reporting all pair-wise comparisons by overloading the interaction term, we prioritize the usability of our results by focusing our analysis on the animation condition that can be combined with a speed encoding choice to optimize a given perceptual task.
The remaining detailed statistics can be found in the supplementary materials. 
\vspace{1mm}

\noindent \textbf{Mean Task:} Generally, encoding speed with mean values of lines produced the highest accuracy on mean comparison tasks.
The performance can be further increased by visualizing data in the \icon{sync_tra_his}\textit{synchronous, trace, history} condition.
On the other hand, we recommend designers to avoid using the \icon{seq_notra_nohis}\textit{sequential, no trace, no history} condition for mean comparison tasks, which led to the worst performance. 
\vspace{1mm}

\noindent \textbf{Variance Task:} Generally, encoding speed with variance values of lines produced the highest accuracy on variance comparison tasks.
Designers can even further increase performance accuracy by visualizing data in the\icon{sync_tra_his}\textit{synchronous, trace, history} condition. 
We recommend designers avoid the \icon{seq_tra_nohis}\textit{sequential, trace, no history} condition for variance comparison tasks as it leads to the worst performance. 
\vspace{1mm}

\noindent \textbf{Outlier Task:} Generally, not using speed encoding for the outlier task would lead to the highest performance.
Therefore, we compared the overall performance on this task when no speed encoding was used. 
The \icon{seq_tra_his}\textit{sequential, trace, history} condition led to the highest performance, while the \icon{seq_notra_nohis}\textit{sequential, no trace, no history} condition led to the worst performance (refer to Experiment 2 for detail). 
\vspace{1mm}

\noindent \textbf{Takeaway:} In the mean comparison task, we found synchronous conditions to be the most beneficial, likely because participants can simultaneously compare the mean values of the lines without having to rely on their working memory, which they have to do in the sequential conditions.
For the variance task, including traces in synchronous lines with history helped, but including traces in sequential lines without history led to the lowest performance. When lines were played out synchronously, it was easier to spot which lines were faster. When synchronous lines preserve histories as well, faster lines have their history on the screen for a longer time, helping people to find out which lines were faster. Even though showing the trace helps people visualize the variance in each line, showing the lines one at a time and without history requires people's working memory of the shape and speed of each line, making it more difficult in comparison.
Similarly, for the outlier task, sequential lines work when trace and history are showing, but yield the lowest performance without trace and history. Without either trace or history, people cannot directly visualize the shape of each line, and also need to memorize the shape of each line to compare across the lines. And especially when lines were played out sequentially without traces, people need to find outlying shapes depicted by multiple moving dots at the same time.

\section{Exp 4 Takeaways and Real World Data}
\label{Exp4}
Finally, we examined how these animation conditions we tested can lead to different viewer takeaways in a more ecologically valid setting, using real-world data from Dolby labs. 

\subsection{Participant, Design, and Procedure}
We recruited 195 participants from Prolific.co~\cite{palan2018prolific}.
After applying the same exclusion criteria as previous experiments,
we were left with 187 individuals (52 female, 132 male, 3 non-binary or self-described, $M_{\textit{age}}$ = 33.72, $SD_{\textit{age}}$ = 10.25).
The study took on average 22 minutes and participants were compensated at the rate of \$12 per hour.

We generated the visualization using 
real-world time-series performance data captured from three Content Delivery Networks (CDN) was shared by Dolby Laboratories.~\cite{dolby} The data contained time-to-first-byte (TTFB) measurements taken at a rate of 900 measurements per hour over a two-week interval in June 2022.
The resulting visualizations are line charts where movements of latency values are shown on the y-axis, time is shown on the x-axis, and each service is represented by a line of a unique color, similar to the experimental set-up in our previous studies. 
We randomly selected two subsets of this data and generated two sets of visualizations (seven visualizations per set, one for each visualization condition), each depicting latency values over time in a different location (A and B).

After consenting to the study, participants first completed a mini VLAT~\cite{pandey2023mini}. 
The average accuracy for all participants is 93.05\%, meaning that the participants overall have high visualization literacy. 
Next, they read a visualization depicting the latency of four services over time in location A in a randomly selected condition out of the seven conditions (sequential/synchronized, yes/no trace, yes/no history, static), and reported the salient message or pattern they saw in a free-response question. 
They then saw the six animation conditions plus the static chart and ranked them based on preference (1 = most preferred, 7 = least preferred). 
All seven visualizations depicted the same data of latency over time in location B. \revision{(preference results reported in Section 4.3)} 
The survey concluded with demographic questions asking about their level of education and experiences working with and designing animated charts \revision{(details see supplementary materials)}, as well as a debrief summarizing the purpose of the study. 

\subsection{Results: \revision{Qualitative Report}}
We coded participants' takeaways following visualization tasks identified from prior work, including Cottam et al.,~\cite{cottam2012watch}, Burns et al.,\cite{burns2020tasks}, Saket et al.\cite{saket2019tasks}, and Shneiderman~\cite{shneiderman2003eyes}. 
The most commonly mentioned tasks were \textit{variance} \revision{(the overall deviation of values from the average)}, \textit{shape} \revision{(the shape of part of the visualization or the entire visualization)}, \textit{retrieving or deriving values} \revision{(directly reading values of the x-axis or y-axis)}, and \textit{outliers} \revision{(outlying values in the visualization)}. \revision{see the supplementary materials for details)}.

The lines continuously fluctuate so that participants were likely to describe or even compare the fluctuations of the lines (considered as both \textit{variance} and \textit{shape}). The significant outliers in the visualizations also stood out, so often mentioned as well (considered as both \textit{outliers} and \textit{shape}). Under unlimited time to examine the visualization, participants were able to extract more detailed information, such as looking at the axes to describe specific numbers that mark interesting patterns in the visualizations (considered as \textit{retrieving or deriving values}).
\vspace{1mm}

\noindent  \textbf{Takeaway:} In addition to perceptual tasks, animation design styles also afford different qualitative takeaways from data, validated by existing work highlighting the value of animation beyond analytic tasks measured by speed and accuracy, like facilitating engagement~\cite{amini2018hooked, szopinski2020visualize}.

\section{Summary Findings and Design Implications}
\label{disc}

\revision{Prior work (e.g. \cite{robertson2008effectiveness}) has shown that people typically prefer animated visualizations over static ones because they are more engaging, but we found people preferred static visualization the most. 
This might be because participants were asked to rank their preferences after performing perceptual tasks, so their preference rankings were based on the effectiveness of these visualizations rather than their engagement.} We observed a consistent discrepancy between the animation styles and speeds that people subjectively preferred and the ones that objectively afford higher performance on perceptual tasks. 
Generally, people disliked the \icon{sync_notra_nohis}\textit{synchronous, no trace, no history} condition, but they tend to perform well with it on perceptual tasks. 

Considering both subjective and objective measures, synchronous displays outperform sequential displays\revision{, consistent with prior results that staggering is not so helpful in animated visualizations \cite{chevalier2014not}}. 
Prior findings show that traces could be confusing, especially for large data sets \cite{kriglstein2012animation}, we found that showing traces could lower performance for evaluating the mean values, despite it being preferred. 
History, which is also subjectively preferred, enhances judgments of variance. 
The \icon{sync_tra_his}\textit{synchronous, trace, history} animation condition had the highest performance for all perceptual tasks, while \icon{seq_notra_nohis}\textit{sequential, no trace, no history} condition had the lowest performance. 

Generally, if a designer has to use sequential displays, preserving histories and traces can improve viewer performance, especially when there is more data.
In terms of speed, people preferred the lines in a sequential display to move approximately 3 times faster than the same data in a synchronous display.  
They also preferred animations with trace to play slower and those with history to play faster.
Motion-based visual encoding can improve performance on perceptual tasks. 

Encoding the mean or variance values of lines with speed can increase performance accuracy on the corresponding task, but encoding outlier information with speed can hurt task performance. 

\section{Limitation and Future Directions}
\label{limitation}
We only tested the animation techniques in line charts with up to four lines. \revision{Future work should test for the generalizability of our findings across other chart types, datasets, and animation styles, including complex dashboards or more information-rich analytic environments.
For example, a monitoring device showing dynamic time series might fade away the history lines without completely removing them. Future work can investigate animation affordances on visual analysis in these more real-world applications.}

Further, we only considered time steps from two to eight seconds for our speed investigations. Future work can increase the range of display time and test for discriminability over small intervals.
Following a more psychophysics approach~\cite{elliott2020design}, for example, future work can determine just-noticeable differences and limits of animation perception, such as the fastest speed for which people can still accurately complete perceptual tasks. 

We also did not account for individual differences systematically.
\revision{We took an online crowd-sourcing approach to balance out individual differences, which can have their own drawbacks~\cite{borgo2017crowdsourcing, borgo2018information}.}
Viewer preference could be swayed by their background~\cite{liu2020survey}, expertise~\cite{xiong2019curse}, and personality~\cite{ottley2015manipulating}.
Participants could also have differed on the amount of cognitive load and attention resources they were willing to commit to our study. 
\revision{With further enhancement to the quality of experimental data collection, such as additional checks for participant engagement filters and study environment,
future work can model individual differences} in perceptual task performance, such as employing methods introduced by Davis et al.~\cite{davis2022risks}. 
Finally, it is possible that participants interpreted the speed encoding as an implicit variable showing the time factor in data.
For example, one interpretation might be that some lines were changing values faster than others because they came from sources with different data processing speeds.

Future work can examine the strategies participants used to gain a deeper understanding of animation perception in visualizations.
This might also inspire additional data encoding channels to afford more efficient data extraction. 

\acknowledgments{
We thank the online participants who made this study possible. We also thank our reviewers for their helpful feedback. This work was partially supported by NSF award IIS-2237585, IIS-2107490, and IIS-2311575, as well as Dolby.
}


\bibliographystyle{abbrv-doi-hyperref}
\bibliography{reference-0630}

\end{document}